\documentclass[sigconf]{acmart}

\AtBeginDocument{%
  \providecommand\BibTeX{{%
    \normalfont B\kern-0.5em{\scshape i\kern-0.25em b}\kern-0.8em\TeX}}}

\copyrightyear{2023} 
\acmYear{2023} 
\setcopyright{acmlicensed}\acmConference[MMSys '23]{Proceedings of the 14th ACM Multimedia Systems Conference}{June 7--10, 2023}{Vancouver, BC, Canada}
\acmBooktitle{Proceedings of the 14th ACM Multimedia Systems Conference (MMSys '23), June 7--10, 2023, Vancouver, BC, Canada}
\acmPrice{15.00}
\acmDOI{10.1145/3587819.3592545}
\acmISBN{979-8-4007-0148-1/23/06}

\usepackage{tabularx}
\usepackage{array}
\usepackage{makecell}
\usepackage{caption}
\usepackage{subcaption}
\usepackage{xcolor}
\newcommand\red[1]{\textcolor{red}{#1}}

\newcommand\green[1]{\textcolor[RGB]{0,100,0}{#1}}

\usepackage[belowskip=2pt,aboveskip=2pt]{caption}

\usepackage{multirow}
\usepackage[group-separator={,},group-minimum-digits=4]{siunitx}

\begin{document}

\title{TotalDefMeme: A Multi-Attribute Meme dataset on \\ Total Defence in Singapore}


\author{    Nirmalendu Prakash}
\authornote{Both authors contributed equally to this research.}
\affiliation{%
  \institution{Singapore University of \\Technology and Design}
  \city{Singapore}
  \country{Singapore}
}
\email{nirmalendu_prakash@sutd.edu.sg}

\author{Ming Shan Hee}
\authornotemark[1]
\affiliation{%
  \institution{Singapore University of \\Technology and Design}
  \city{Singapore}
  \country{Singapore}
}
\email{mingshan_hee@mymail.sutd.edu.sg}

\author{Roy Ka-Wei Lee}
\affiliation{%
  \institution{Singapore University of \\Technology and Design}
  \city{Singapore}
  \country{Singapore}
}
\email{roy_lee@sutd.edu.sg}



\renewcommand{\shortauthors}{Prakash et al.}

\begin{abstract}
\textit{Total Defence} is a defence policy combining and extending the concept of military defence and civil defence. While several countries have adopted total defence as their defence policy, very few studies have investigated its effectiveness. With the rapid proliferation of social media and digitalisation, many social studies have been focused on investigating policy effectiveness through specially curated surveys and questionnaires either through digital media or traditional forms. However, such references may not truly reflect the underlying sentiments about the target policies or initiatives of interest. People are more likely to express their sentiment using communication mediums such as starting topic thread on forums or sharing memes on social media. Using Singapore as a case reference, this study aims to address this research gap by proposing \textsf{TotalDefMeme}, a large-scale multi-modal and multi-attribute meme dataset that captures public sentiments toward Singapore's Total Defence policy. Besides supporting social informatics and public policy analysis of the Total Defence policy, \textsf{TotalDefMeme} can also support many downstream multi-modal machine learning tasks, such as aspect-based stance classification and multi-modal meme clustering. We perform baseline machine learning experiments on \textsf{TotalDefMeme} and evaluate its technical validity, and present possible future interdisciplinary research directions and application scenarios using the dataset as a baseline.
\end{abstract}



\maketitle


\section{Introduction}
\label{sec:int
roduction}
The growing popularity of social media platforms has led to the popularisation of internet memes, which provide the means for self-expression, connection, and social influence. Memes, typically presented as images with short superimposed text, are often designed to be relatable, humorous, and shareable. While memes have slowly become a fundamental aspect of contemporary digital culture, where most of the memes are humorous and meant for comic relief, there are increasingly more that are meant to express grievances and negative sentiments towards public individuals and institutions. Soh~\cite{WeeYangSoh20202digital} highlights memes as a strategy for pragmatic resistance, allowing users to evade charges of sedition due to the authorship dispersal nature of memes. This meant that it was possible to disseminate information under the masquerade of humour, and craft narratives to shape public sentiments without much consideration of repercussions. Therefore, memes are a good communication medium for investigating the public's sentiment toward a country's government and its policies. 

An example of such a policy may be encompassed under a country's defence strategy. \textit{Total Defence} is a defence policy combining and extending the concept of military defence and civil defence. Several countries, including Sweden, Switzerland, Russia, Ukraine, Singapore, etc., have adopted the policy framework. It entails a high level of readiness of the state and its society to defend itself in catastrophic events such as war, crisis, and natural disasters. For instance, Singapore's \textit{Total Defence} strategy focuses on six pillars in preparing and assessing the countries' readiness for catastrophic events: \textit{military}, \textit{civil}, \textit{economic}, \textit{social}, \textit{psychological}, and \textit{digital}. While several countries have adopted the \textit{Total Defence} framework as a policy guidance, very few studies have investigated its effectiveness through the ground perceptions of the masses apart from specially curated surveys which may not reflect the underlying sentiments of society. This is primarily due to the unavailability of datasets to facilitate such an analysis.  
\begin{figure}
    \centering
    \includegraphics[scale=0.8]{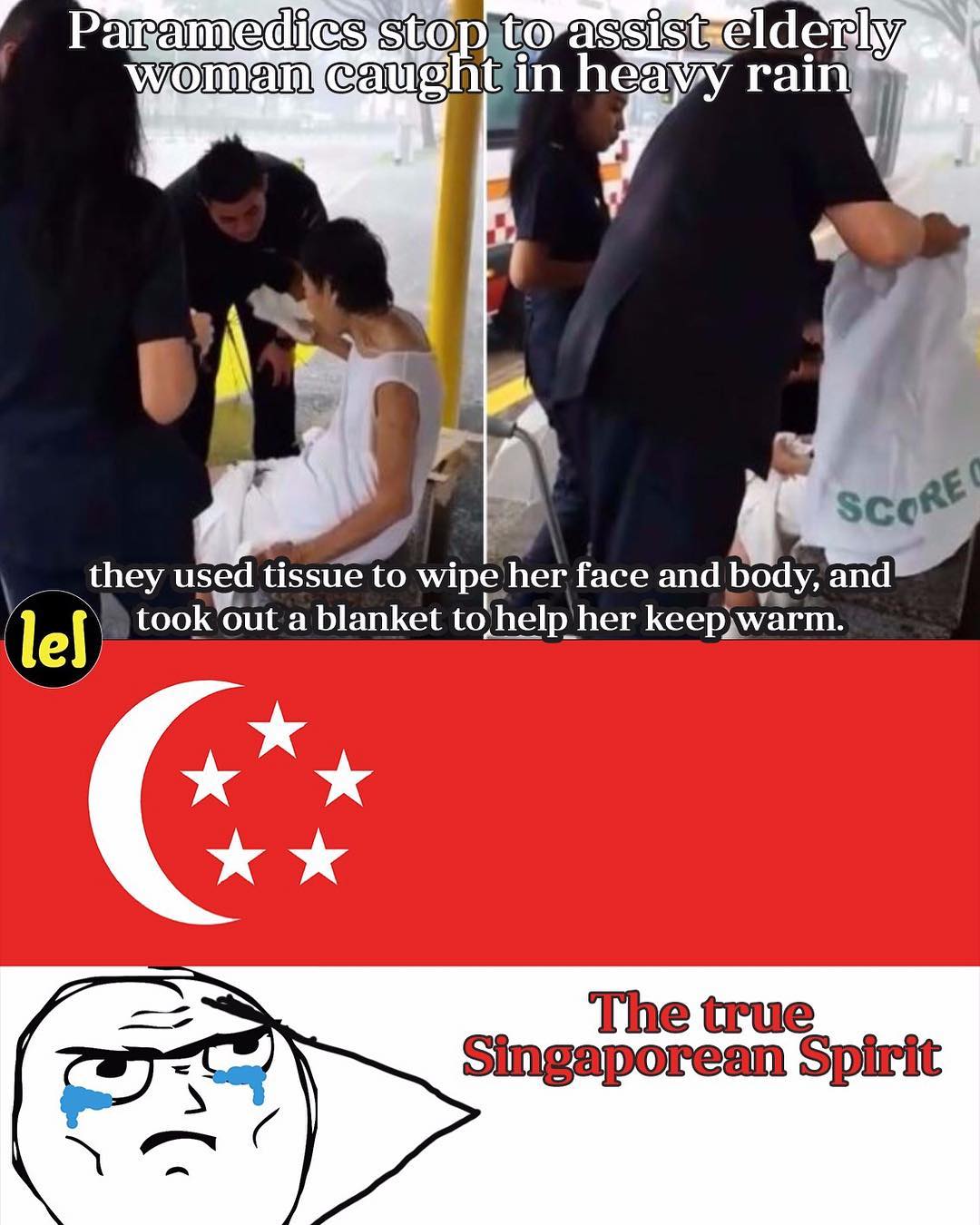}
    \caption{An example of a \textsf{TotalDefMeme} meme and its multi-attribute annotation. The meme praises Singapore's Civil Defence by showing the paramedics going beyond their work duties to help an elderly woman.}
    \label{fig:example}
\end{figure}

This study aims to fill the data limitation gap by constructing \textsf{TotalDefMeme}\footnote{\url {https://gitlab.com/bottle_shop/meme/TotalDefMemes}} \footnote{*This dataset should only be used for non-commercial research purposes. Should you find any meme in violation of copyright, please notify us and we will take it down.}, a large-scale multimodal and multi-attribute meme dataset that captures public sentiments toward Singapore's total defence policy. Figure \ref{fig:example} illustrates an example meme in our \textsf{TotalDefMeme} dataset. The meme depicts paramedics doing a good deed beyond their work duties, expressing stances that are supportive of Singapore's civil defence, which is of the country's six total defence pillars.



Besides supporting social informatics and public policy analysis of total defence policy, \textsf{TotalDefMeme} can also support many downstream multimodal machine learning tasks. Existing meme research studies have extensively studied memes with malicious intent by analyzing and releasing large datasets on a few binary attributes, such as hate speech \cite{kiela2020hateful,mathias-etal-2021-findings}, harmfulness \cite{pramanick2021detecting}, misinformation \cite{qu2022disinfomeme}, offensiveness \cite{suryawanshi-etal-2020-multimodal}, information diffusion \cite{Leskovec2009MemetrackingAT} etc. However, there has been limited focus on understanding the topics and stances expressed in memes. Furthermore, most of the existing meme datasets focused on the English language and western cultural context. In contrast, our \textsf{TotalDefMeme} is a multimodal meme dataset based on Southeast Asia and Singapore context, where the language captured is predominantly colloquial with a contextualized variant of English, otherwise known as Singlish\footnote{\url{ https://en.wikipedia.org/wiki/Singlish}}. Our \textsf{TotalDefMeme} dataset also aims to address the existing meme research gaps by providing a dataset with multiple attributes that could facilitate downstream machine learning tasks such as multimodal aspect-based stance analysis and multimodal meme clustering.

 

We summarize our contributions as follows:

\begin{itemize}
    \item We construct \textsf{TotalDefMeme}, a large-scale multimodal and multi-attribute meme dataset that captures public sentiments toward Singapore's total defence policy.
   \item We discuss and suggest possible future interdisciplinary research directions and application scenarios using the dataset.
   \item We perform a set of baseline machine learning experiments on \textsf{TotalDefMeme} and evaluate its technical validity. 
\end{itemize}


\section{Related Works}
\label{sec:related}
\textbf{Meme Analysis.} Multimodal study of memes has gained traction in recent years, which has resulted in multiple datasets with different objectives. Recently Facebook released a Hateful Memes Challenge\cite{kiela2020hateful}, where the task is to classify memes as hateful and non-hateful. Mathias et el.\cite{mathias-etal-2021-findings} further extended the Hateful Memes Challenge dataset by adding the protected category (i.e., \textit{race}, \textit{disability}, \textit{religion}, \textit{nationality}, \textit{sex}) that has been attacked by the meme, and the type of attack (i.e., \textit{contempt}, \textit{mocking}, \textit{inferiority}, \textit{slurs}, \textit{exclusion}, \textit{dehumanizing}, \textit{inciting violence}). These hateful meme datasets have encouraged researchers to developed hateful meme detection solutions~\cite{zhu2022multimodal,hee2022explaining,cao2023prompting,lee2021disentangling}. Suryavanshi et al.\cite{suryawanshi-etal-2020-multimodal} constructed a meme dataset on the 2016 United States presidential elections with offensiveness labels. Similarly, Pramanik et al. \cite{pramanick2021detecting} released a dataset called \textit{HarMeme}, to study harmful memes. The researchers retrieved memes by querying Google Image Search with keywords related to covid-19. As part of SemEval 2020 tasks, Sharma et al.\cite{sharma-etal-2020-semeval} released a dataset called Memotion, to capture fine-grained emotions expressed in memes. Pramanick et al. \cite{Pramanick2021ExerciseIT} further proposed a multi-hop attention-based deep learning approach to leverage spatial-domain correspondence between visual and textual modalities to extract fine-grained feature representations for sentiment classification on the Memotion dataset. In a recent study, Qu et al.\cite{qu2022disinfomeme} examined memes that spread misinformation. They collect memes from Reddit on three topics, namely, \textit{Covid-19}, \textit{BLM}, and \textit{veganism}, and annotate binary misinformation labels. While more meme datasets are available to support downstream multimodal machine learning tasks, most are annotated with simple binary labels and can only support limited classification tasks. Furthermore, most existing memes datasets only contain English memes based on western cultures. 

Our study value-adds to the existing meme studies by providing a multimodal meme dataset that captures multiple attributes to support more multimodal machine learning tasks. Specifically, our proposed \textsf{TotalDefMeme} dataset contains memes annotated with multi-attribute labels, such as the types of memes, the topic discussed, the total defence pillars affected, and the stance towards the pillars. Furthermore, \textsf{TotalDefMeme} contains memes on Southeast Asia and Singapore's cultural context, where the language captured is \textit{Singlish}. The multi-attribute nature of our \textsf{TotalDefMeme} dataset also facilitates machine learning tasks, such as aspect-based stance analysis and multimodal topic clustering, that existing datasets could not support.

\begin{figure*}
    \centering
    \includegraphics[scale=0.8]{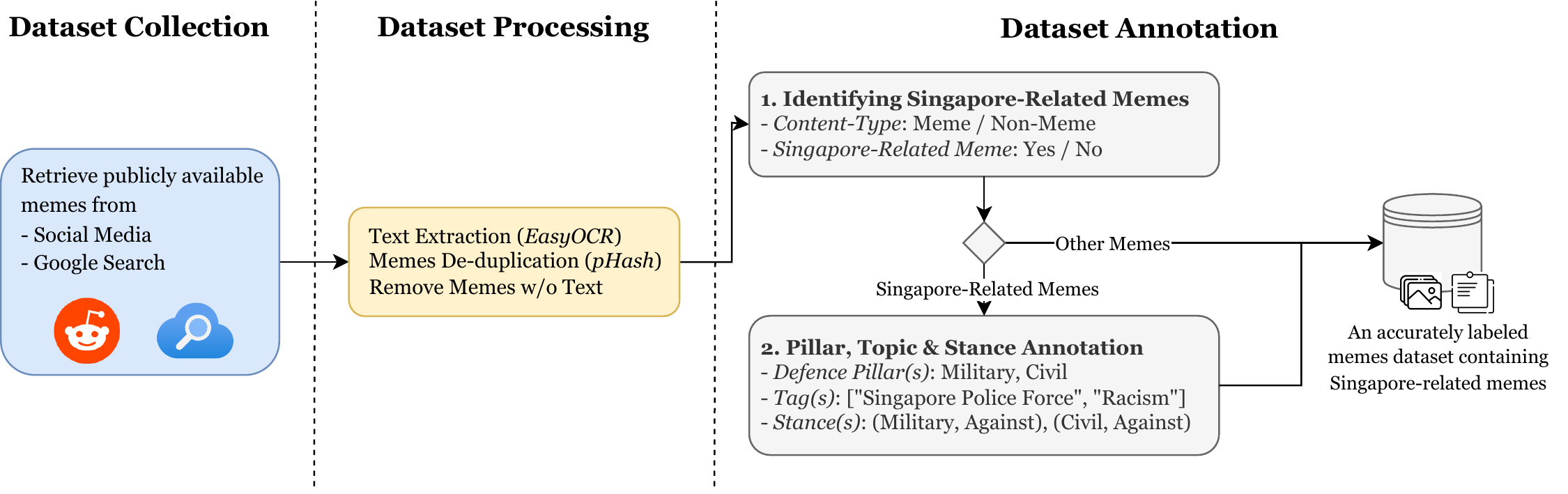}
    \caption{Data Collection \& Annotation Pipeline.} 
    \label{fig:data-annotation-pipeline}
\end{figure*}

\textbf{Total Defence.} The Total Defence policy is a defence policy that incorporates both military and civil defence strategies. It is adopted by countries such as Finland, Norway, Sweden, and Singapore. The policy emphasizes a high level of readiness against any potential dangers and catastrophes, including war, crisis, and natural disasters, for the state and its society. Hence, as a key defence policy, Total Defence serves as the primary guide to the needs of the government and its citizens.

While various countries employ the Total Defence policy, their implementation is not globally uniform. Our research involved collecting and analyzing memes pertaining to Singapore's Total Defence policy, which comprises six pillars: \textit{Military}, \textit{Civil}, \textit{Economic}, \textit{Social}, \textit{Psychological}, and \textit{Digital}. Through our examination of multiple sources\footnote{https://www.mindef.gov.sg/oms/imindef/mindef\_websites/topics/totaldefence/about.html}\footnote{https://www.scdf.gov.sg/home/community-volunteers/community-preparedness/total-defence}, we have summarized the definitions of each pillar as follows:

\begin{itemize}

\item \textbf{Military Defence}: Strong and formidable defence force made up of Regulars and National Servicemen, and supported by the entire nation
\item \textbf{Civil Defence}: Collective effort of the society to spot signs of threats, respond effectively and recover quickly from crisis
\item \textbf{Economic Defence}: Strong and resilient economy that is globally competitive and able to bounce back from any crisis.
\item \textbf{Social Defence}: Bonds that unite us, built on trust and understanding among people of different races and religions, living in harmony and looking out for one another
\item \textbf{Psychological Defence}: The will and resolve to defend our way of life and interests, the fighting spirit to overcome challenges together
\item \textbf{Digital Defence}: Being secure, alert and responsible online.
\end{itemize}


\section{Dataset Construction}
\label{sec:dataset}
In this section, we describe the data annotation pipeline used for constructing the \textsf{TotalDefMeme} dataset and provide a preliminary analysis of the dataset. The pipeline, shown in Figure \ref{fig:data-annotation-pipeline}, mainly comprises three phases: \textit{dataset collection}, \textit{dataset processing}, and \textit{dataset annotation}. 

\subsection{Dataset Collection}
To collect memes related to Singapore’s Total Defence, we adopted a keyword-based approach using Google Search to obtain more relevant memes. We studied the Total Defence concepts across multiple sources and crafted the appropriate keywords such as ``\textit{police force}'', ``\textit{racism}'', ``\textit{phone scams}''. Subsequently, these keywords are inserted into a template query: “Singapore <keyword> Memes”. We further scraped various publicly available groups on popular social media platforms, such as Reddit and Instagram, to increase our coverage\footnote{Note that we did not use content from any private or restricted pages}. Including memes from social media platforms enables our examination of recent and viral Singapore-related memes. We obtained a dataset of 7,200 diverse memes through these methods.

\subsection{Dataset Processing}
To align with our research on multimodal memes, we applied strict filtering criteria. First, we performed a simple filtering on the quality of the memes where we removed memes with image resolution smaller than 224x224 pixels and memes with text exceeding 50-word tokens. Second, we applied a text extraction tool using the EasyOCR\footnote{https://www.jaided.ai/easyocr/} algorithm to retrieve the text found within the meme and remove the ones without text. Third, we utilized the pHash algorithm to identify groups of duplicates and preserved each group's memes with the highest resolution. Finally, we retain 5,301 memes, which serves as the final \textsf{TotalDefMeme} dataset.


\begin{figure}
    \includegraphics[scale=0.21]{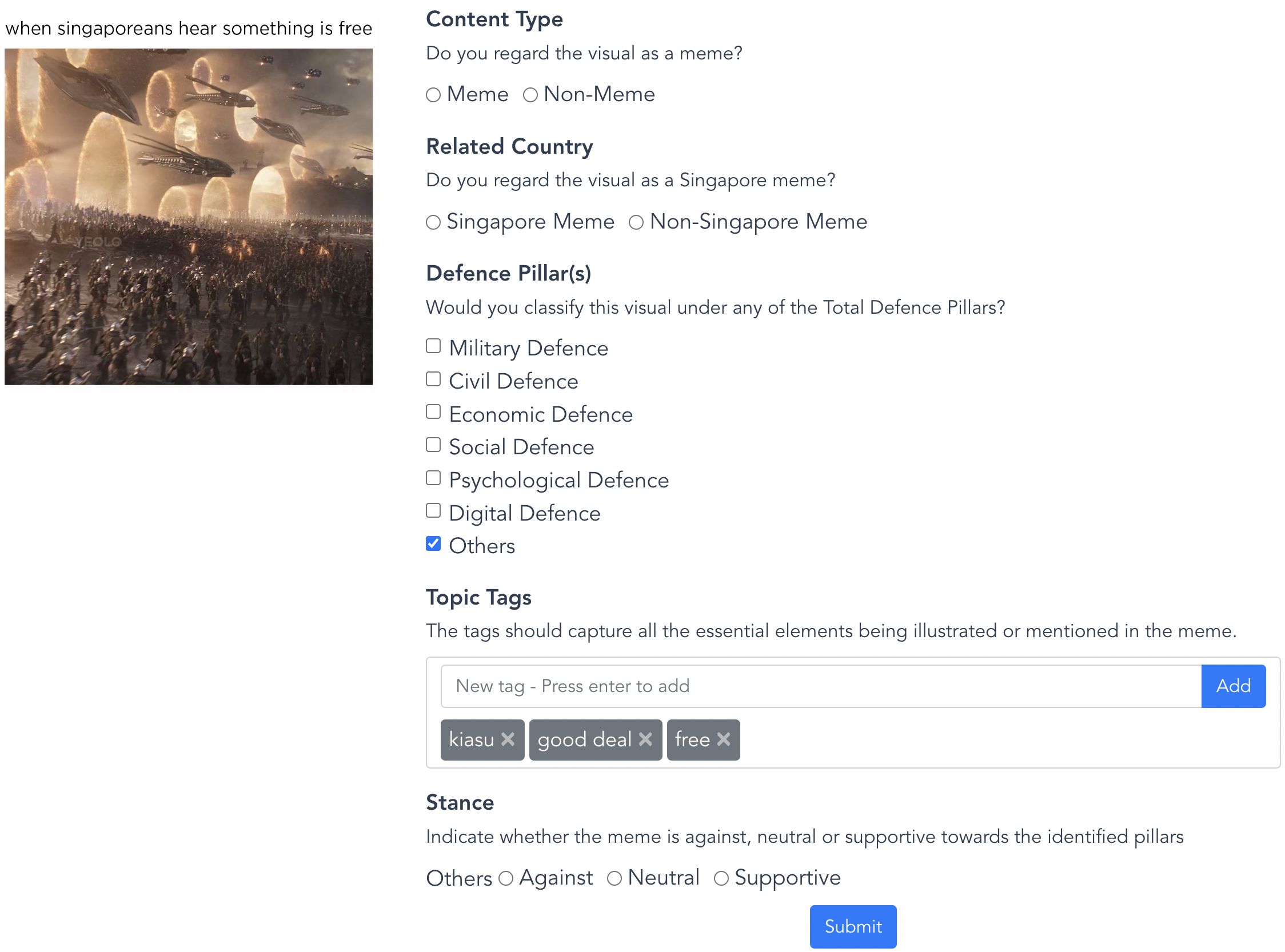}
    \caption{Snapshot of the user interface used for \textit{Meme Identification} and \textit{Pillars, Topics \& Stances annotation}}
    \label{fig:annotation-platform}
\end{figure}

\begin{figure*}
    \includegraphics[scale=0.5]{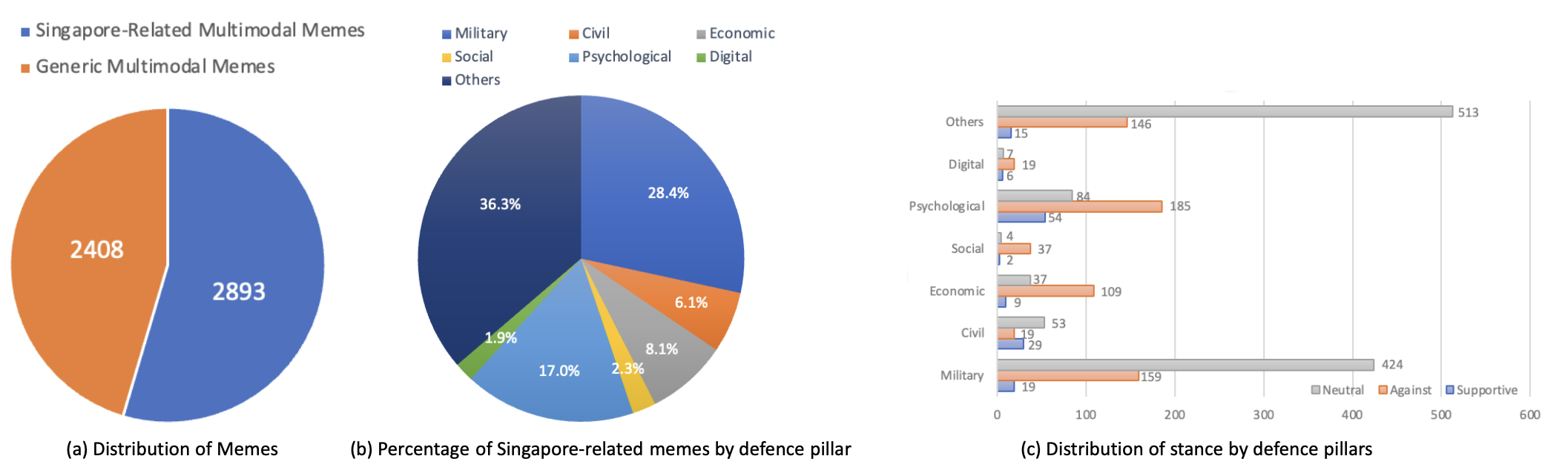}
    \caption{Statistics about \textsf{TotalDefMeme} dataset}
    \label{fig:dataset_statistic}
\end{figure*}

\subsection{Dataset Annotation}
We recruited six annotators who are familiar with Singapore's culture and knowledgeable about Singapore's Total Defence concept from a pool of shortlisted Singaporean undergraduates who have passed a short screening survey. The annotators are tasked with identifying the Singapore-related memes and annotating the pillars, topics, and stances of these Singapore-related memes. Screenshots of the user interface can be seen in Figure \ref{fig:annotation-platform}

\textbf{Meme Region Identification} The annotators will conduct a manual review and distinguish Singapore-related multi-modal memes from the diverse collection of visuals, including infographics and posters. Each meme can be categorised into two types: generic memes and Singapore-related memes. For example, a meme that addresses the topic of working overtime shall be considered a generic meme, as this scenario can occur in the majority of countries, while a meme that addresses the Singapore Police Force shall be considered a Singapore-related meme.


\begin{table}[t]
\centering
    \caption{Krippendorff's Alpha for each task: Meme Type (3 classes), Defence Pillar (7 classes) and Stance (3 classes).}
\begin{tabular}{c c c c }
    \hline
     Types & Pillars & Stances \\ 
     \hline\hline
     0.65 & 0.55 & 0.21 \\    
     \hline
    \end{tabular}

\label{tab:table-kappa}
\end{table}


\textbf{Pillars, Topics \& Stances Annotation} Upon identifying the Singapore-related meme, the annotators annotate the pillars, topics, and stances of the meme. The annotators will first assign the memes' defence pillars: \textit{military}, \textit{civil}, \textit{economic}, \textit{social}, \textit{psychological}, \textit{digital}, or \textit{others}. We included the ``\textit{others}'' pillar as an option as there are memes that talk about daily occurrences and modern affairs, which do not involve any of the six defence pillars. Next, they annotate the relevant topic tags associated with the meme (i.e., nouns, pronouns, and phrases) in a free-text format. The annotators can enter the most appropriate tags and are encouraged to enter as many relevant tags as possible. Lastly, the annotators annotate the meme's stances towards the assigned pillars: \textit{support}, \textit{against}, or \textit{neutral}.

\textbf{Quality Control Measures} To ensure the reliability of the dataset, each meme is annotated by two annotators. If the disagreements contain similar opinions, the overlap annotations will be considered correct labels. However, if there are disagreements with entirely different perspectives, a third annotator will be brought in to provide an additional annotation for the meme. The overlapping annotations between at least two annotators will then be considered the correct labels. In the extreme case where all three annotators have different opinions, the meme will be flagged and removed from the dataset. Finally, we held review discussions with the annotators to discuss the annotations with disagreements, allowing our annotators to receive feedback and improve their annotations.

\subsection{Dataset Analysis}
Figure~\ref{fig:dataset_statistic} shows the statistical summary of the \textsf{TotalDefMeme} dataset. More than half of the memes in our dataset are Singapore-related. On the defence pillar annotation, we notice the class label imbalance. \textit{Military} is the dominant class, with close to a third of the memes (36\%) labeled to be related to military defence. Conversely, only 1.9\% and 2.3\% of the memes are labeled as \textit{Digital} and \textit{Social} classes, respectively. Interestingly, we also noted a substantial number of memes annotated as \textit{Others}, which captures other Singapore-related but non total defence related topics, adding diversity to the \textsf{TotalDefMeme} dataset.  

We computed Krippendorff's Alpha score to measure the inter-annotator agreement for the various annotation labels. Table \ref{tab:table-kappa} presents the scores for memes' \textit{types}, \textit{pillars}, and \textit{stances}. Although the task requires in-depth contextual knowledge of Singapore and total defence, the annotators have achieved a moderate agreement with alpha scores of 0.65 and 0.55 for \textit{types} and \textit{pillars}, respectively, indicating quality annotations. The agreement for \textit{stances} is weaker (alpha score of 0.21) due to the subjectivity of the task. Nevertheless, we will release the annotations and annotators' id in our released dataset to facilitate further analysis.

\section{Interdisciplinary Research and Applications}
\label{sec:applications}
The purpose of publishing this unique dataset is to motivate social science and computer science researchers to explore interdisciplinary research and propose novel machine learning tasks. As a starting point, we foresee this dataset to have several applications and usage scenarios. A few examples are as follows:
\begin{itemize}
    \item Analysis and assessment of country's total defence readiness
    \item Multimodal aspect-based stance classification
    \item Multimodal meme clustering
    \item Domain adaption of meme analysis
\end{itemize}

\begin{figure*}
    \centering
    \includegraphics[scale=0.35]{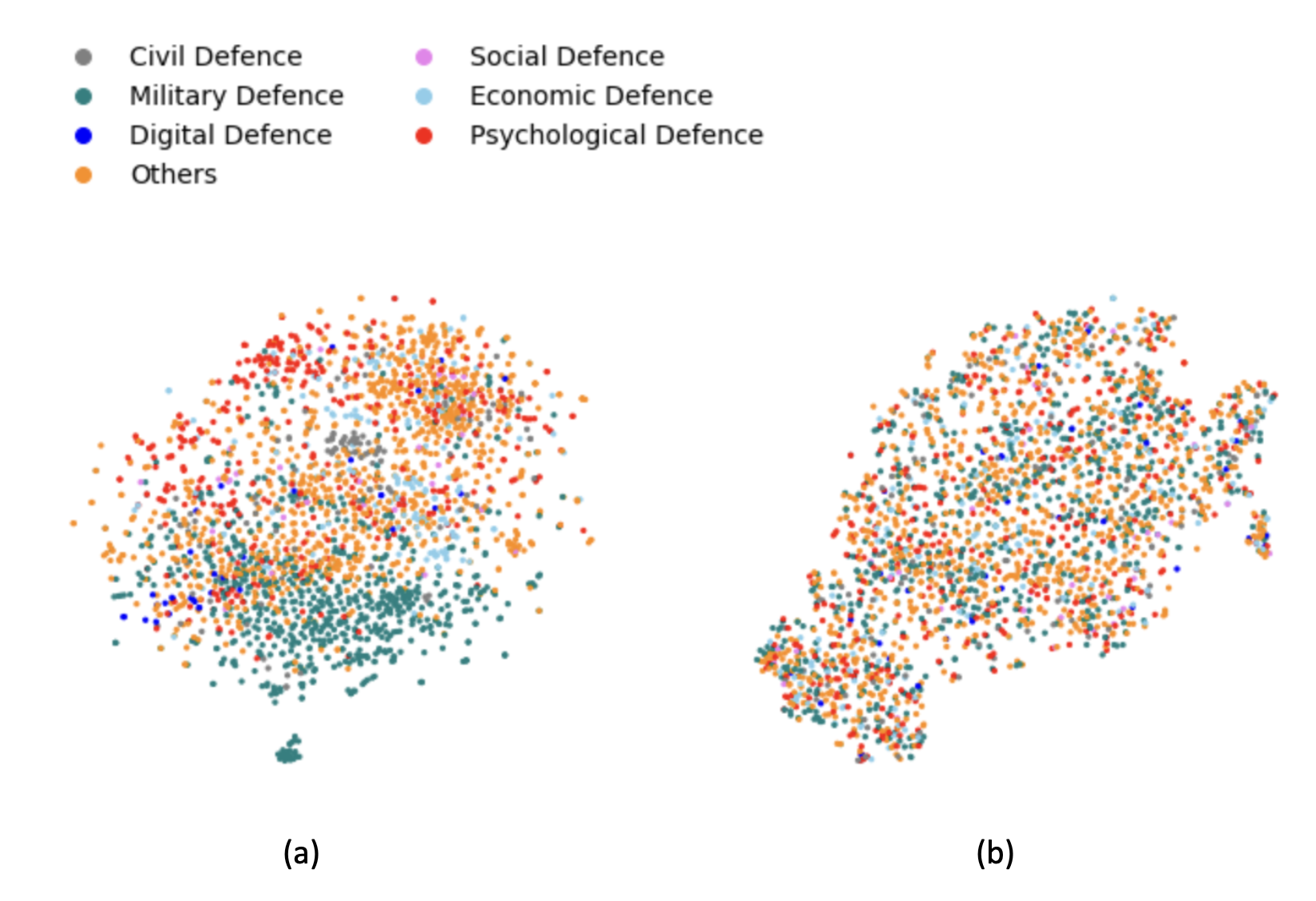}
    \caption{t-SNE visualization of (a) CLIP and (b) VisualBERT embeddings.}
    \label{fig:tsne_plot}
\end{figure*}

A clear use case for the \textsf{TotalDefMeme} dataset is to support interdisciplinary computational social science research. Computational social scientists can analyze the dataset to answer social informatics and policy-related questions. Furthermore, the annotation framework can also be applied to annotate and analyze total defence memes from other countries, such as Russia, Ukraine, etc.  

Existing studies have explored sentiment and emotion classification in memes~\cite{sharma-etal-2020-semeval}. Our proposed dataset extends this line of research to support multimodal aspect-based stance classification, where the task aims to predict the stance towards an entity, concept, or events illustrated in a meme. Specifically, in \textsf{TotalDefMeme}, a possible task is to predict the stance towards the total defence pillar illustrated in a given meme. This is challenging as the machine learning model will need to interpret multi-modality information in the meme to predict both the total defence pillars and the corresponding stances. We will benchmark several baselines on this task using \textsf{TotalDefMeme} next Section~\ref{sec:experiments}.

Existing studies have explored performing clustering of memes using unimodal approaches \cite{ferrara2013clustering,jafariasbagh2014clustering,dang2015visual,10.1145/3278532.3278550,dubey2018memesequencer, theisen2020automatic}, where the memes' visuals are first captioned with text before applying topic models to cluster the memes or perform clustering on the memes' visuals representations. However, such approaches neglect the rich information in the interaction between textual and visual modality. Furthermore, evaluating the clustering algorithms is challenging as the ground truth is often unavailable. We hope that \textsf{TotalDefMeme} can encourage researchers to propose multimodal meme clustering methods. \textsf{TotalDefMeme} memes are also annotated with the topics and total defence pillars, which could serve as ground truth for evaluating the multimodal meme clustering algorithms. We will also benchmark several baselines on this task using \textsf{TotalDefMeme}. 

\begin{table}[t]
    \centering
      \caption{Silhoutte Score \& NMI of the k-Mean clusters using various embeddings.}
    \begin{tabular}{ c | c | c }
     \hline
    Embedding & Silhouette Score & NMI\\
     \hline\hline
    BERT  & 0.030  & 0.020\\
    VGG16  & -0.013 & 0.010\\
    VisualBERT  & \textbf{0.094} & 0.006 \\
    CLIP  & 0.012  & \textbf{0.091}\\
    \hline
    \end{tabular}
    \label{tab:clustering_metric}
\end{table}

Another potential use case of \textsf{TotalDefMeme} is to support domain adaption of meme analysis. Multimodal meme analysis techniques can be trained using \textsf{TotalDefMeme} and evaluate its performance on other meme datasets. This is particularly useful for analyzing total defence memes collected from other countries.

\section{Experiments}
\label{sec:experiments}

\subsection{Meme Clustering}
To perform the clustering of memes, we first compute the memes' representation using various unimodal and multimodal models. Specifically, for unimodal models, we adopted pre-trained BERT~\cite{devlin2018bert} to compute a meme's representation using its texts, and VGG16~\cite{simonyan2014very} to extract a meme's representation using its visual. For multimodal models, we use VisualBERT~\cite{li2019visualbert} to compute the meme's representation using both its visual and text. We also utilize CLIP~\cite{radford2021learning} to extract a meme's visual and text representations separately before concatenating them to get the final meme's representation.

\begin{table}[t]
    \centering
    \caption{Accuracy scores for pillar and stance classification tasks.}
    \begin{tabular}{ c | c | c }
     \hline
     Models & Pillar & Stance \\
     \hline\hline
    BERT & 0.18 & 0.30 \\
    VGG & 0.14 & 0.30 \\
    VisualBERT & 0.05 & 0.22 \\
    CLIP & \textbf{0.57} & \textbf{0.54} \\
    \hline
    \end{tabular}

    \label{tab:pillars_stance_f1}
\end{table}

\newcolumntype{M}{>{\centering\arraybackslash}m{2.2cm}}
\newcolumntype{N}{>{\centering\arraybackslash}m{3.2cm}}
\begin{table*}[ht]
  \small
  \centering
  \begin{tabular}{M|N|N|N|N}
  \hline
    \multirow{2}{*}{\textbf{Meme}} & \begin{minipage}[!b]{0.40\columnwidth}
		\centering
		\raisebox{-.7\height}{\includegraphics[width=\linewidth]{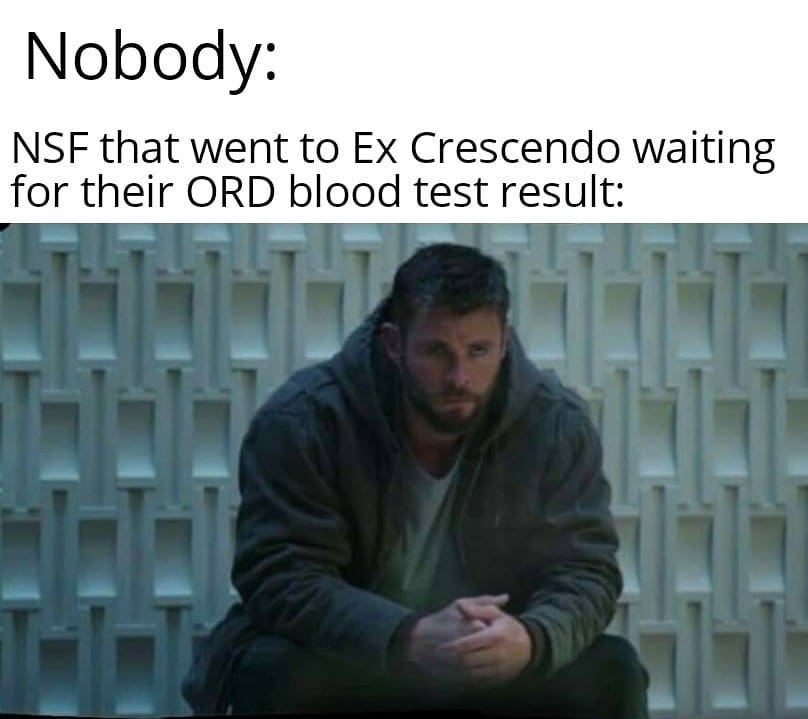}}
	\end{minipage} &
    \begin{minipage}[!b]{0.40\columnwidth}
		\centering
		\raisebox{-.5\height}
  {\includegraphics[width=\linewidth]{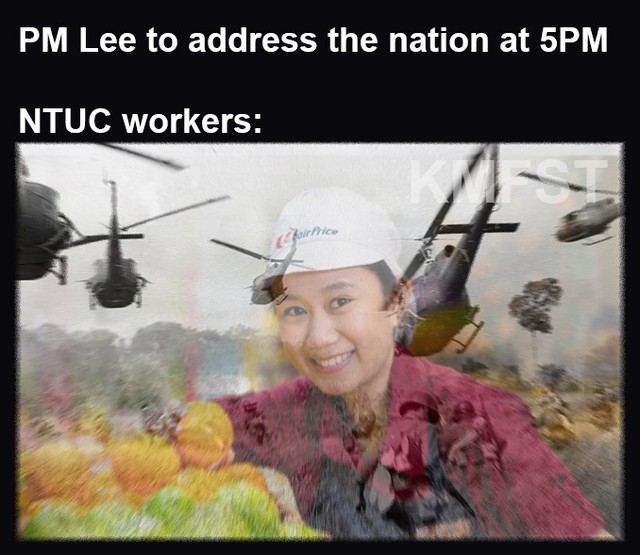}}
	\end{minipage} &
    \begin{minipage}[!b]{0.40\columnwidth}
		\centering
		\raisebox{-.5\height}{\includegraphics[width=\linewidth]{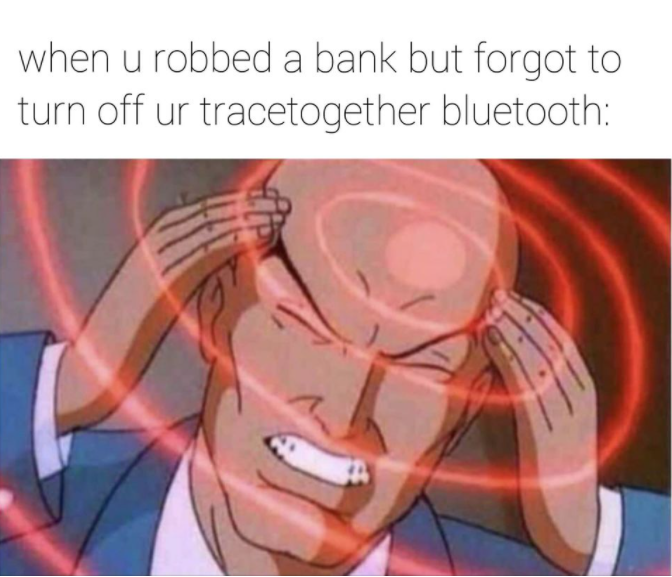}}
	\end{minipage} & 
    \begin{minipage}[!b]{0.35\columnwidth}
		\centering
		\raisebox{-.5\height}{\includegraphics[width=\linewidth]{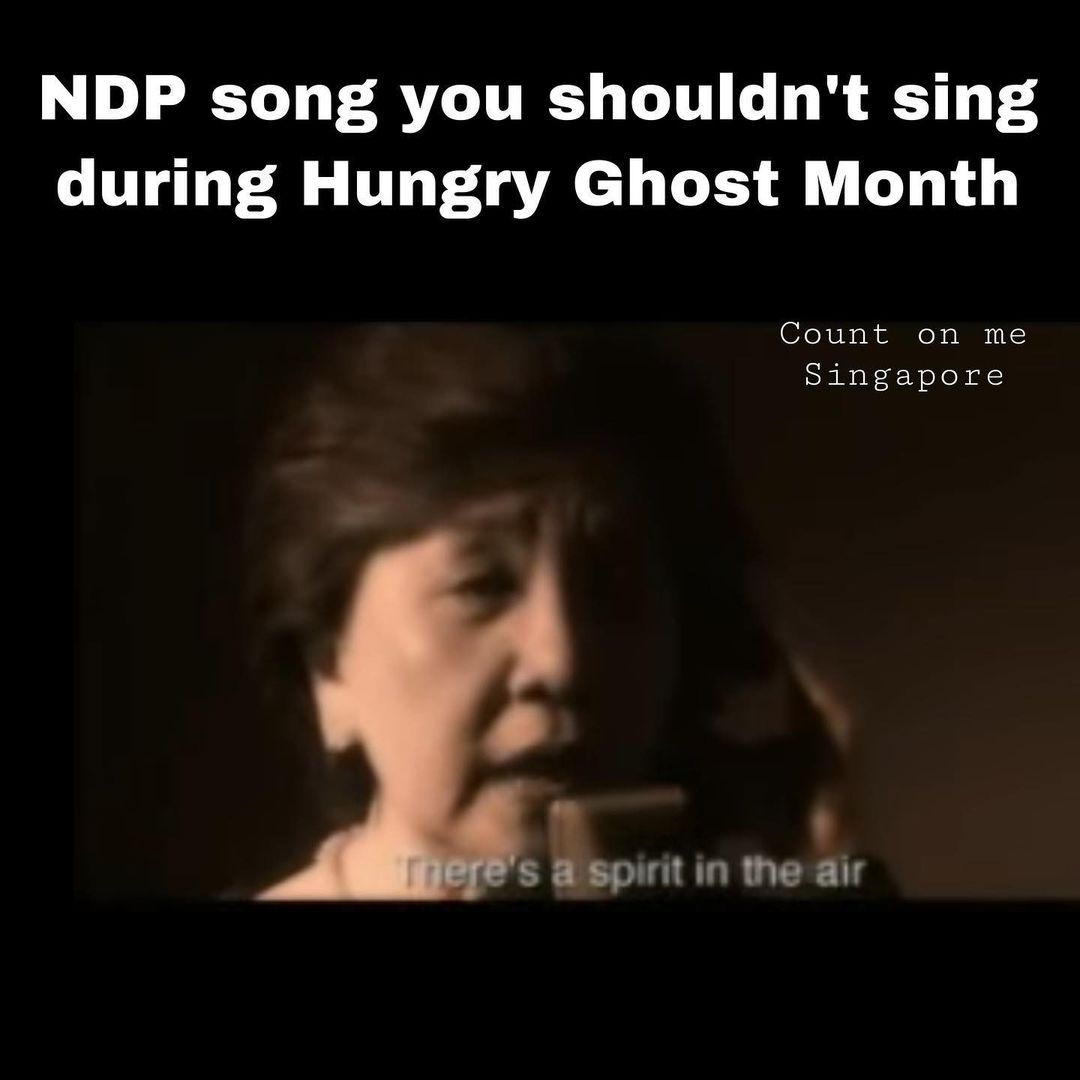}}
	\end{minipage}\\
    & (a) & (b) & (c) & (d) \\
    \hline
    \textbf{Predicted Pillar(s)} & \green{Military} & \red{Psychological} & \green{Digital} & \green{Psychological} \\\hline    \textbf{Ground-Truth Pillar(s)} & Military  & Economic & Digital & Psychological \\
    \hline
    \textbf{Predicted Stance(s)} & \green{Military - Neutral} & \red{Psychological - Against} & \green{Digital - Against} & \red{Psychological - Against} \\\hline
    \textbf{Ground-Truth Stance(s)} & Military - Neutral & Economic - Against & Digital - Against & Psychological - Neutral \\\hline
    \end{tabular}
    
  \caption{Example of CLIP pillars and stances classification results. Correct and incorrect predictions are color coded as \green{green} and \red{red}, respectively.}
  \label{tab:sample_results}
\end{table*}

The computed meme's representations are subsequently input into k-mean clustering algorithm. We set the cluster size to 7, aiming to obtain pillar-wise semantic clusters. Table~\ref{tab:clustering_metric} shows the evaluation of the k-mean clustering results using the various unimodal and multimodal embeddings. Specifically, we computed the Silhouette and  Normalized Mutual Information(NMI) scores for learned clusters. We observe that using the multimodal embeddings have achieved better clustering performance; VisualBERT embeddings has the highest Silhouette score (0.094), while CLIP embeddings achieved the highest NMI (0.091). Conversely, using only the text (i.e., BERT) or visual (i.e., VGG16) embeddings results in poor clustering performance. 

To further examine the multimodal embeddings, we show the t-SNE visualization of the k-mean clustering algorithm that was trained using CLIP and VisualBERT embeddings in Figure~\ref{fig:tsne_plot}. The scatterplots are color-coded with the annotated total defence pillars. Interestingly, we observed that the CLIP embeddings seem to be clustering together according to the total defence pillar. In contrast, the VisualBERT embeddings seem to be scattered with no clear patterns based on pillars. Nevertheless, we note that the Silhouette score and NMI of simple k-mean clustering with multimodal embeddings are still very low, highlighting the challenge of this task and the need to develop better multimodal meme clustering methods.

\subsection{Aspect-based Stance Classification}
To perform aspect-based stance classification, we first stratified split \textsf{TotalDefMeme} by pillar into train, validation, and test sets using the ratio 60/20/20. Next, we train the baselines using the train set. For unimodal baselines, we obtain BERT~\cite{devlin2018bert} embeddings using the memes' text, and VGG16~\cite{simonyan2014very} embeddings using the memes' visual, and train a MLP layer on top of the embeddings. For multimodal baselines, we use VisualBERT~\cite{li2019visualbert} and  CLIP~\cite{radford2021learning} embeddings and train a MLP layer using both the memes' text and visual. As \textsf{TotalDefMeme} is an imbalance dataset, we upsampled the minority classes in the training set when training the model.

For this experiment, we formulate the aspect-based stance classification as a multitask classification problem where the model predicts both the pillars and pillars' stances as separate outputs. To achieve this, we modified the loss function in the linear output layer of the baseline classifiers to predict the memes' pillars and corresponding pillars' stances. Table~\ref{tab:pillars_stance_f1} shows baselines' accuracy scores for the pillar and stance classification tasks. We observe that the CLIP achieved the best performance on both tasks. 

To further examine the predictions, we show example predictions of the best-performing model, CLIP, in Table~\ref{tab:sample_results}. In example memes (a) and (c), CLIP has correctly predicted the pillars and stances. More interestingly, we also noted that there are cases where the models made incorrect predictions in this task. For instance, in example meme (b), CLIP correctly predicted the ``psychology'' pillar but missed out on the ``economic pillar''. For example meme (d), CLIP correctly predicted the pillar but predicted an incorrect stance.  The case studies and experimental results suggest room for improvement in the aspect-based stance classification task. We hope the release of \textsf{TotalDefMeme} can encourage more researchers to develop better aspect-based stance classification techniques.

\section{Conclusion}
\label{sec:conclusion}
In this paper, we have introduced \textsf{TotalDefMeme}, a novel multimodal multi-attribute meme dataset based on \textit{Total Defence}. The dataset extends beyond the traditional hateful/offensive meme datasets by incorporating an aspect of public policy analysis from a social community perspective. We believe that social media memes capture a raw take on the bread and butter issues relating to governmental policies and initiatives, thereby providing the underlying public sentiments that may not be accurately captured from public surveys and questionnaires. This dataset is collected from the web without any preset limit on the number or domain of topics, and is annotated with the relevant defence pillars, associated topics and stance labels based on expert knowledge. Therefore, it provides the research community with a baseline for a comprehensive study on social media memes with respect to Total Defence or related governmental policies. We envisage the dataset to continue to push the boundaries for multimodal meme understanding and further the research for visual-language aspect-based stance classification and multimodal meme clustering.

\bibliographystyle{ACM-Reference-Format}
\bibliography{ref}


\begin{thebibliography}{23}


\ifx \showCODEN    \undefined \def \showCODEN     #1{\unskip}     \fi
\ifx \showDOI      \undefined \def \showDOI       #1{#1}\fi
\ifx \showISBNx    \undefined \def \showISBNx     #1{\unskip}     \fi
\ifx \showISBNxiii \undefined \def \showISBNxiii  #1{\unskip}     \fi
\ifx \showISSN     \undefined \def \showISSN      #1{\unskip}     \fi
\ifx \showLCCN     \undefined \def \showLCCN      #1{\unskip}     \fi
\ifx \shownote     \undefined \def \shownote      #1{#1}          \fi
\ifx \showarticletitle \undefined \def \showarticletitle #1{#1}   \fi
\ifx \showURL      \undefined \def \showURL       {\relax}        \fi
\providecommand\bibfield[2]{#2}
\providecommand\bibinfo[2]{#2}
\providecommand\natexlab[1]{#1}
\providecommand\showeprint[2][]{arXiv:#2}

\bibitem[\protect\citeauthoryear{Cao, Lee, Chong, and Jiang}{Cao
  et~al\mbox{.}}{2023}]%
        {cao2023prompting}
\bibfield{author}{\bibinfo{person}{Rui Cao}, \bibinfo{person}{Roy Ka-Wei Lee},
  \bibinfo{person}{Wen-Haw Chong}, {and} \bibinfo{person}{Jing Jiang}.}
  \bibinfo{year}{2023}\natexlab{}.
\newblock \showarticletitle{Prompting for Multimodal Hateful Meme
  Classification}.
\newblock \bibinfo{journal}{\emph{arXiv preprint arXiv:2302.04156}}
  (\bibinfo{year}{2023}).
\newblock


\bibitem[\protect\citeauthoryear{Dang, Moh'd, Gruzd, Milios, and Minghim}{Dang
  et~al\mbox{.}}{2015}]%
        {dang2015visual}
\bibfield{author}{\bibinfo{person}{Anh Dang}, \bibinfo{person}{Abidalrahman
  Moh'd}, \bibinfo{person}{Anatoliy Gruzd}, \bibinfo{person}{Evangelos Milios},
  {and} \bibinfo{person}{Rosane Minghim}.} \bibinfo{year}{2015}\natexlab{}.
\newblock \showarticletitle{A visual framework for clustering memes in social
  media}. In \bibinfo{booktitle}{\emph{2015 IEEE/ACM International Conference
  on Advances in Social Networks Analysis and Mining (ASONAM)}}. IEEE,
  \bibinfo{pages}{713--720}.
\newblock


\bibitem[\protect\citeauthoryear{Devlin, Chang, Lee, and Toutanova}{Devlin
  et~al\mbox{.}}{2018}]%
        {devlin2018bert}
\bibfield{author}{\bibinfo{person}{Jacob Devlin}, \bibinfo{person}{Ming-Wei
  Chang}, \bibinfo{person}{Kenton Lee}, {and} \bibinfo{person}{Kristina
  Toutanova}.} \bibinfo{year}{2018}\natexlab{}.
\newblock \showarticletitle{Bert: Pre-training of deep bidirectional
  transformers for language understanding}.
\newblock \bibinfo{journal}{\emph{arXiv preprint arXiv:1810.04805}}
  (\bibinfo{year}{2018}).
\newblock


\bibitem[\protect\citeauthoryear{Dubey, Moro, Cebrian, and Rahwan}{Dubey
  et~al\mbox{.}}{2018}]%
        {dubey2018memesequencer}
\bibfield{author}{\bibinfo{person}{Abhimanyu Dubey}, \bibinfo{person}{Esteban
  Moro}, \bibinfo{person}{Manuel Cebrian}, {and} \bibinfo{person}{Iyad
  Rahwan}.} \bibinfo{year}{2018}\natexlab{}.
\newblock \showarticletitle{Memesequencer: Sparse matching for embedding image
  macros}. In \bibinfo{booktitle}{\emph{Proceedings of the 2018 World Wide Web
  Conference}}. \bibinfo{pages}{1225--1235}.
\newblock


\bibitem[\protect\citeauthoryear{Ferrara, JafariAsbagh, Varol, Qazvinian,
  Menczer, and Flammini}{Ferrara et~al\mbox{.}}{2013}]%
        {ferrara2013clustering}
\bibfield{author}{\bibinfo{person}{Emilio Ferrara}, \bibinfo{person}{Mohsen
  JafariAsbagh}, \bibinfo{person}{Onur Varol}, \bibinfo{person}{Vahed
  Qazvinian}, \bibinfo{person}{Filippo Menczer}, {and}
  \bibinfo{person}{Alessandro Flammini}.} \bibinfo{year}{2013}\natexlab{}.
\newblock \showarticletitle{Clustering memes in social media}. In
  \bibinfo{booktitle}{\emph{2013 IEEE/ACM International Conference on Advances
  in Social Networks Analysis and Mining (ASONAM 2013)}}. IEEE,
  \bibinfo{pages}{548--555}.
\newblock


\bibitem[\protect\citeauthoryear{Hee, Lee, and Chong}{Hee
  et~al\mbox{.}}{2022}]%
        {hee2022explaining}
\bibfield{author}{\bibinfo{person}{Ming~Shan Hee}, \bibinfo{person}{Roy Ka-Wei
  Lee}, {and} \bibinfo{person}{Wen-Haw Chong}.}
  \bibinfo{year}{2022}\natexlab{}.
\newblock \showarticletitle{On Explaining Multimodal Hateful Meme Detection
  Models}. In \bibinfo{booktitle}{\emph{Proceedings of the ACM Web Conference
  2022}}. \bibinfo{pages}{3651--3655}.
\newblock


\bibitem[\protect\citeauthoryear{JafariAsbagh, Ferrara, Varol, Menczer, and
  Flammini}{JafariAsbagh et~al\mbox{.}}{2014}]%
        {jafariasbagh2014clustering}
\bibfield{author}{\bibinfo{person}{Mohsen JafariAsbagh},
  \bibinfo{person}{Emilio Ferrara}, \bibinfo{person}{Onur Varol},
  \bibinfo{person}{Filippo Menczer}, {and} \bibinfo{person}{Alessandro
  Flammini}.} \bibinfo{year}{2014}\natexlab{}.
\newblock \showarticletitle{Clustering memes in social media streams}.
\newblock \bibinfo{journal}{\emph{Social Network Analysis and Mining}}
  \bibinfo{volume}{4}, \bibinfo{number}{1} (\bibinfo{year}{2014}),
  \bibinfo{pages}{1--13}.
\newblock


\bibitem[\protect\citeauthoryear{Kiela, Firooz, Mohan, Goswami, Singh,
  Ringshia, and Testuggine}{Kiela et~al\mbox{.}}{2020}]%
        {kiela2020hateful}
\bibfield{author}{\bibinfo{person}{Douwe Kiela}, \bibinfo{person}{Hamed
  Firooz}, \bibinfo{person}{Aravind Mohan}, \bibinfo{person}{Vedanuj Goswami},
  \bibinfo{person}{Amanpreet Singh}, \bibinfo{person}{Pratik Ringshia}, {and}
  \bibinfo{person}{Davide Testuggine}.} \bibinfo{year}{2020}\natexlab{}.
\newblock \showarticletitle{The hateful memes challenge: Detecting hate speech
  in multimodal memes}.
\newblock \bibinfo{journal}{\emph{Advances in Neural Information Processing
  Systems}}  \bibinfo{volume}{33} (\bibinfo{year}{2020}),
  \bibinfo{pages}{2611--2624}.
\newblock


\bibitem[\protect\citeauthoryear{Lee, Cao, Fan, Jiang, and Chong}{Lee
  et~al\mbox{.}}{2021}]%
        {lee2021disentangling}
\bibfield{author}{\bibinfo{person}{Roy Ka-Wei Lee}, \bibinfo{person}{Rui Cao},
  \bibinfo{person}{Ziqing Fan}, \bibinfo{person}{Jing Jiang}, {and}
  \bibinfo{person}{Wen-Haw Chong}.} \bibinfo{year}{2021}\natexlab{}.
\newblock \showarticletitle{Disentangling hate in online memes}. In
  \bibinfo{booktitle}{\emph{Proceedings of the 29th ACM International
  Conference on Multimedia}}. \bibinfo{pages}{5138--5147}.
\newblock


\bibitem[\protect\citeauthoryear{Leskovec, Backstrom, and Kleinberg}{Leskovec
  et~al\mbox{.}}{2009}]%
        {Leskovec2009MemetrackingAT}
\bibfield{author}{\bibinfo{person}{Jure Leskovec}, \bibinfo{person}{Lars
  Backstrom}, {and} \bibinfo{person}{Jon~M. Kleinberg}.}
  \bibinfo{year}{2009}\natexlab{}.
\newblock \showarticletitle{Meme-tracking and the dynamics of the news cycle}.
  In \bibinfo{booktitle}{\emph{Knowledge Discovery and Data Mining}}.
\newblock


\bibitem[\protect\citeauthoryear{Li, Yatskar, Yin, Hsieh, and Chang}{Li
  et~al\mbox{.}}{2019}]%
        {li2019visualbert}
\bibfield{author}{\bibinfo{person}{Liunian~Harold Li}, \bibinfo{person}{Mark
  Yatskar}, \bibinfo{person}{Da Yin}, \bibinfo{person}{Cho-Jui Hsieh}, {and}
  \bibinfo{person}{Kai-Wei Chang}.} \bibinfo{year}{2019}\natexlab{}.
\newblock \showarticletitle{Visualbert: A simple and performant baseline for
  vision and language}.
\newblock \bibinfo{journal}{\emph{arXiv preprint arXiv:1908.03557}}
  (\bibinfo{year}{2019}).
\newblock


\bibitem[\protect\citeauthoryear{Mathias, Nie, Mostafazadeh~Davani, Kiela,
  Prabhakaran, Vidgen, and Waseem}{Mathias et~al\mbox{.}}{2021}]%
        {mathias-etal-2021-findings}
\bibfield{author}{\bibinfo{person}{Lambert Mathias}, \bibinfo{person}{Shaoliang
  Nie}, \bibinfo{person}{Aida Mostafazadeh~Davani}, \bibinfo{person}{Douwe
  Kiela}, \bibinfo{person}{Vinodkumar Prabhakaran}, \bibinfo{person}{Bertie
  Vidgen}, {and} \bibinfo{person}{Zeerak Waseem}.}
  \bibinfo{year}{2021}\natexlab{}.
\newblock \showarticletitle{Findings of the {WOAH} 5 Shared Task on Fine
  Grained Hateful Memes Detection}. In \bibinfo{booktitle}{\emph{Proceedings of
  the 5th Workshop on Online Abuse and Harms (WOAH 2021)}}.
  \bibinfo{publisher}{Association for Computational Linguistics},
  \bibinfo{address}{Online}, \bibinfo{pages}{201--206}.
\newblock
\urldef\tempurl%
\url{https://doi.org/10.18653/v1/2021.woah-1.21}
\showDOI{\tempurl}


\bibitem[\protect\citeauthoryear{Pramanick, Akhtar, and Chakraborty}{Pramanick
  et~al\mbox{.}}{2021a}]%
        {Pramanick2021ExerciseIT}
\bibfield{author}{\bibinfo{person}{Shraman Pramanick},
  \bibinfo{person}{Md.~Shad Akhtar}, {and} \bibinfo{person}{Tanmoy
  Chakraborty}.} \bibinfo{year}{2021}\natexlab{a}.
\newblock \showarticletitle{Exercise? I thought you said 'Extra Fries':
  Leveraging Sentence Demarcations and Multi-hop Attention for Meme Affect
  Analysis}.
\newblock \bibinfo{journal}{\emph{ArXiv}}  \bibinfo{volume}{abs/2103.12377}
  (\bibinfo{year}{2021}).
\newblock


\bibitem[\protect\citeauthoryear{Pramanick, Dimitrov, Mukherjee, Sharma,
  Akhtar, Nakov, Chakraborty, et~al\mbox{.}}{Pramanick et~al\mbox{.}}{2021b}]%
        {pramanick2021detecting}
\bibfield{author}{\bibinfo{person}{Shraman Pramanick}, \bibinfo{person}{Dimitar
  Dimitrov}, \bibinfo{person}{Rituparna Mukherjee}, \bibinfo{person}{Shivam
  Sharma}, \bibinfo{person}{Md Akhtar}, \bibinfo{person}{Preslav Nakov},
  \bibinfo{person}{Tanmoy Chakraborty}, {et~al\mbox{.}}}
  \bibinfo{year}{2021}\natexlab{b}.
\newblock \showarticletitle{Detecting harmful memes and their targets}.
\newblock \bibinfo{journal}{\emph{arXiv preprint arXiv:2110.00413}}
  (\bibinfo{year}{2021}).
\newblock


\bibitem[\protect\citeauthoryear{Qu, Li, Zhao, Dev, and Chang}{Qu
  et~al\mbox{.}}{2022}]%
        {qu2022disinfomeme}
\bibfield{author}{\bibinfo{person}{Jingnong Qu},
  \bibinfo{person}{Liunian~Harold Li}, \bibinfo{person}{Jieyu Zhao},
  \bibinfo{person}{Sunipa Dev}, {and} \bibinfo{person}{Kai-Wei Chang}.}
  \bibinfo{year}{2022}\natexlab{}.
\newblock \showarticletitle{DisinfoMeme: A Multimodal Dataset for Detecting
  Meme Intentionally Spreading Out Disinformation}.
\newblock \bibinfo{journal}{\emph{arXiv preprint arXiv:2205.12617}}
  (\bibinfo{year}{2022}).
\newblock


\bibitem[\protect\citeauthoryear{Radford, Kim, Hallacy, Ramesh, Goh, Agarwal,
  Sastry, Askell, Mishkin, Clark, et~al\mbox{.}}{Radford et~al\mbox{.}}{2021}]%
        {radford2021learning}
\bibfield{author}{\bibinfo{person}{Alec Radford}, \bibinfo{person}{Jong~Wook
  Kim}, \bibinfo{person}{Chris Hallacy}, \bibinfo{person}{Aditya Ramesh},
  \bibinfo{person}{Gabriel Goh}, \bibinfo{person}{Sandhini Agarwal},
  \bibinfo{person}{Girish Sastry}, \bibinfo{person}{Amanda Askell},
  \bibinfo{person}{Pamela Mishkin}, \bibinfo{person}{Jack Clark},
  {et~al\mbox{.}}} \bibinfo{year}{2021}\natexlab{}.
\newblock \showarticletitle{Learning transferable visual models from natural
  language supervision}. In \bibinfo{booktitle}{\emph{International Conference
  on Machine Learning}}. PMLR, \bibinfo{pages}{8748--8763}.
\newblock


\bibitem[\protect\citeauthoryear{Sharma, Bhageria, Scott, PYKL, Das,
  Chakraborty, Pulabaigari, and Gamb{\"a}ck}{Sharma et~al\mbox{.}}{2020}]%
        {sharma-etal-2020-semeval}
\bibfield{author}{\bibinfo{person}{Chhavi Sharma}, \bibinfo{person}{Deepesh
  Bhageria}, \bibinfo{person}{William Scott}, \bibinfo{person}{Srinivas PYKL},
  \bibinfo{person}{Amitava Das}, \bibinfo{person}{Tanmoy Chakraborty},
  \bibinfo{person}{Viswanath Pulabaigari}, {and} \bibinfo{person}{Bj{\"o}rn
  Gamb{\"a}ck}.} \bibinfo{year}{2020}\natexlab{}.
\newblock \showarticletitle{{S}em{E}val-2020 Task 8: Memotion Analysis- the
  Visuo-Lingual Metaphor!}. In \bibinfo{booktitle}{\emph{Proceedings of the
  Fourteenth Workshop on Semantic Evaluation}}.
  \bibinfo{publisher}{International Committee for Computational Linguistics},
  \bibinfo{address}{Barcelona (online)}, \bibinfo{pages}{759--773}.
\newblock
\urldef\tempurl%
\url{https://doi.org/10.18653/v1/2020.semeval-1.99}
\showDOI{\tempurl}


\bibitem[\protect\citeauthoryear{Simonyan and Zisserman}{Simonyan and
  Zisserman}{2014}]%
        {simonyan2014very}
\bibfield{author}{\bibinfo{person}{Karen Simonyan} {and}
  \bibinfo{person}{Andrew Zisserman}.} \bibinfo{year}{2014}\natexlab{}.
\newblock \showarticletitle{Very deep convolutional networks for large-scale
  image recognition}.
\newblock \bibinfo{journal}{\emph{arXiv preprint arXiv:1409.1556}}
  (\bibinfo{year}{2014}).
\newblock


\bibitem[\protect\citeauthoryear{Soh}{Soh}{2020}]%
        {WeeYangSoh20202digital}
\bibfield{author}{\bibinfo{person}{Wee~Yang Soh}.}
  \bibinfo{year}{2020}\natexlab{}.
\newblock \showarticletitle{Digital protest in Singapore: the pragmatics of
  political Internet memes}. In \bibinfo{booktitle}{\emph{Sage Journals -
  Media, Culture \& Society}}, Vol.~\bibinfo{volume}{42}.
  \bibinfo{pages}{1115--1132}.
\newblock
Issue 7-8.


\bibitem[\protect\citeauthoryear{Suryawanshi, Chakravarthi, Arcan, and
  Buitelaar}{Suryawanshi et~al\mbox{.}}{2020}]%
        {suryawanshi-etal-2020-multimodal}
\bibfield{author}{\bibinfo{person}{Shardul Suryawanshi},
  \bibinfo{person}{Bharathi~Raja Chakravarthi}, \bibinfo{person}{Mihael Arcan},
  {and} \bibinfo{person}{Paul Buitelaar}.} \bibinfo{year}{2020}\natexlab{}.
\newblock \showarticletitle{Multimodal Meme Dataset ({M}ulti{OFF}) for
  Identifying Offensive Content in Image and Text}. In
  \bibinfo{booktitle}{\emph{Proceedings of the Second Workshop on Trolling,
  Aggression and Cyberbullying}}. \bibinfo{publisher}{European Language
  Resources Association (ELRA)}, \bibinfo{address}{Marseille, France},
  \bibinfo{pages}{32--41}.
\newblock
\showISBNx{979-10-95546-56-6}
\urldef\tempurl%
\url{https://aclanthology.org/2020.trac-1.6}
\showURL{%
\tempurl}


\bibitem[\protect\citeauthoryear{Theisen, Brogan, Thomas, Moreira, Phoa,
  Weninger, and Scheirer}{Theisen et~al\mbox{.}}{2020}]%
        {theisen2020automatic}
\bibfield{author}{\bibinfo{person}{William Theisen}, \bibinfo{person}{Joel
  Brogan}, \bibinfo{person}{Pamela~Bilo Thomas}, \bibinfo{person}{Daniel
  Moreira}, \bibinfo{person}{Pascal Phoa}, \bibinfo{person}{Tim Weninger},
  {and} \bibinfo{person}{Walter Scheirer}.} \bibinfo{year}{2020}\natexlab{}.
\newblock \showarticletitle{Automatic Discovery of Meme Genres with Diverse
  Appearances}.
\newblock \bibinfo{journal}{\emph{Association for the Advancement of Artificial
  Intelligence}} (\bibinfo{year}{2020}), \bibinfo{pages}{1--13}.
\newblock


\bibitem[\protect\citeauthoryear{Zannettou, Caulfield, Blackburn,
  De~Cristofaro, Sirivianos, Stringhini, and Suarez-Tangil}{Zannettou
  et~al\mbox{.}}{2018}]%
        {10.1145/3278532.3278550}
\bibfield{author}{\bibinfo{person}{Savvas Zannettou}, \bibinfo{person}{Tristan
  Caulfield}, \bibinfo{person}{Jeremy Blackburn}, \bibinfo{person}{Emiliano
  De~Cristofaro}, \bibinfo{person}{Michael Sirivianos},
  \bibinfo{person}{Gianluca Stringhini}, {and} \bibinfo{person}{Guillermo
  Suarez-Tangil}.} \bibinfo{year}{2018}\natexlab{}.
\newblock \showarticletitle{On the Origins of Memes by Means of Fringe Web
  Communities}. In \bibinfo{booktitle}{\emph{Proceedings of the Internet
  Measurement Conference 2018}} (Boston, MA, USA) \emph{(\bibinfo{series}{IMC
  '18})}. \bibinfo{publisher}{Association for Computing Machinery},
  \bibinfo{address}{New York, NY, USA}, \bibinfo{pages}{188–202}.
\newblock
\showISBNx{9781450356190}
\urldef\tempurl%
\url{https://doi.org/10.1145/3278532.3278550}
\showDOI{\tempurl}


\bibitem[\protect\citeauthoryear{Zhu, Lee, and Chong}{Zhu
  et~al\mbox{.}}{2022}]%
        {zhu2022multimodal}
\bibfield{author}{\bibinfo{person}{Jiawen Zhu}, \bibinfo{person}{Roy Ka-Wei
  Lee}, {and} \bibinfo{person}{Wen~Haw Chong}.}
  \bibinfo{year}{2022}\natexlab{}.
\newblock \showarticletitle{Multimodal zero-shot hateful meme detection}. In
  \bibinfo{booktitle}{\emph{14th ACM Web Science Conference 2022}}.
  \bibinfo{pages}{382--389}.
\newblock


\end{thebibliography}

\appendix


\end{document}